\documentclass[aps,prl,onecolumn,floatfix,tightenlines,showpacs,longbibliography,notitlepage]{revtex4-1}
\usepackage{amsfonts}
\usepackage{color}
\usepackage{amsmath}    
\usepackage{epsfig}
\usepackage{graphicx}
\usepackage{dcolumn}
\usepackage{bm}
\usepackage[breaklinks,colorlinks = true,linkcolor = blue,urlcolor=blue,citecolor=blue]{hyperref}

\begin{document}

\title{Uncertainty Growth in Stably Stratified Turbulence}
\author{Mrinal Jyoti Powdel}
 \email{mrinal.jyoti@icts.res.in}
\affiliation{International Centre for Theoretical Sciences, Tata Institute of Fundamental Research, Bengaluru 560089, India}
\author{Samriddhi Sankar Ray}%
 \email{samriddhisankarray@gmail.com}
\affiliation{International Centre for Theoretical Sciences, Tata Institute of Fundamental Research, Bengaluru 560089, India}

\begin{abstract}
We investigate uncertainty growth and chaotic dynamics in statistically steady,
stably stratified three-dimensional turbulence. Using direct numerical
simulations of the Boussinesq equations, we quantify the divergence of
initially infinitesimal perturbations via twin simulations and decorrelator
diagnostics. At short times, perturbations exhibit exponential growth, allowing
	us to define a (largest) Lyapunov exponent. We systematically examine how this
exponent depends on stratification strength, quantified by the
Brunt--V\"{a}is\"{a}l\"{a} frequency and the Froude number, in a parameter
regime relevant to oceanic flows. We find that increasing stratification leads
to a monotonic reduction of the Lyapunov exponent, indicating suppressed
chaoticity. Despite this reduction, uncertainty growth retains the universal
temporal sequence observed in homogeneous isotropic turbulence---initial decay,
exponential growth, and saturation. The growth phase is characterized by
self-similar decorrelator spectra, but exhibits strong anisotropy: uncertainty
spreads much more slowly along the stratification direction than horizontally,
with the disparity increasing with stratification strength. An analysis of the
decorrelator evolution equation reveals that the suppression of chaos arises
primarily from strain-mediated alignment dynamics rather than direct buoyancy
coupling. Our results provide a quantitative characterization of predictability
and uncertainty growth in stratified turbulence and highlight the utility of
decorrelator-based methods for anisotropic geophysical flows.  
\end{abstract}

\maketitle

\section{Introduction}

Predictability in turbulent flows is a central problem in fluid mechanics, with
direct relevance to weather forecasting, climate modelling, oceanic transport,
and environmental dispersion. Even for the idealised case of homogeneous and
isotropic turbulence, the quantitative characterisation of sensitivity to
infinitesimal perturbations remains incomplete. Classical arguments due to
Ruelle~\cite{Ruelle} linked the largest Lyapunov exponent $\lambda$ to Kolmogorov
phenomenology, predicting a scaling with the smallest dynamically active time
scale $\tau_\eta$.  Subsequent developments incorporating intermittency
corrections and multifractal models~\cite{Frisch-Parisi,Frisch_1995} refined these ideas~\cite{Crisanti_multifractal}. Advances in direct
numerical simulations (DNSs) over the past two decades have enabled
increasingly precise measurements of Lyapunov exponents in homogeneous and
isotropic turbulence, yet discrepancies between theoretical predictions, model
systems, and numerical data persist, underlining the subtle nature of chaos in
turbulent flows~\cite{Becetal2006,Siddhartha,Mohan_higher,Boffetta2017,BereraPRL,Ray2018,Vassilicos_2023,Ge_Rolland_Vassilicos_2025}.

A useful recent approach~\cite{Lightcone} to probing sensitivity in high-dimensional dynamical
systems is to track the time evolution of the difference between two initially
nearly identical states evolved under the same dynamics. In turbulent flows,
this naturally leads to the study of such \textit{decorrelators}, which
quantify how velocity or scalar fields diverge in time and space. Such
diagnostics provide a direct, non-perturbative measure of uncertainty growth
and allow one to extract Lyapunov exponents as well as study the spatial
structure and scale-by-scale propagation of perturbations. Recent applications
of this framework to fully developed turbulence~\cite{banerjee2025intermittent} and similar nonlinear
hydrodynamic equations~\cite{GTEuler,mukherjee2023intermittency} have revealed a robust dynamical sequence: An initial
decay of perturbations followed by a period of exponential growth with a
self-similar spectrum and eventual saturation. Remarkably, this appears to be
universal for three-dimensional turbulence~\cite{banerjee2025intermittent}.

While homogeneous and isotropic, fully developed turbulence provides a
fundamental reference state, most natural flows of geophysical relevance differ
markedly from this idealisation. The oceans and atmosphere are often strongly
influenced by stable density stratification, which introduces buoyancy forces,
supports internal gravity waves through the Brunt--V\"ais\"al\"a frequency $N$,
and imposes a preferred vertical direction~\cite{Davidson_2013, rileyannurev, staquetannurev, caulfield_annurev, Caulfield_open_question}. Stratified turbulence is therefore
intrinsically anisotropic and arises from the nonlinear interaction of vortical
motions and wave dynamics. Decades of theoretical, numerical and experimental
work have established that stratification suppresses vertical motions at low
Froude numbers, promotes the formation of layered structures, and alters energy
transfer pathways relative to homogeneous and isotropic turbulence~\cite{Linden01011979, Riley1981, Fernando_1988, Itsweire_Helland_1989, Ruddick1989597, Park_Whitehead_Gnanadeskian_1994, Nicolleau_Vassilicos_2000, Chomaz_self_similarity, Dubrulle_transition,  Lindborg_2006, Lindborg_Brethouwer_2008, Augier_Galtier_Billant_2012, Kimura_Herring_2012, rorai_mininni_largescale, Herbert_Marino_Rosenberg_Pouquet_2016,glazunov2019layered, Maffioli_sign, Labarre2024, Varanasi2025}. While these
effects are now well understood at the level of flow structure and energetics,
their consequences for the chaotic dynamics of turbulence---including the rate
of uncertainty growth, the spreading of perturbations in space, and the overall
predictability of stratified flows---have received comparatively little
attention.

Predictability and sensitivity to initial conditions in stratified flows have
been explored in several settings, but predominantly for decaying or rotational
turbulence. In such studies~\cite{Diaz2024}, predictability is
typically characterised using relative error growth or predictability times in
decaying and not statistically stationary dynamics. Lagrangian approaches
\cite{peng2024applications, hariri2022analysis,
bettencourt2013characterization} based on finite-time Lyapunov exponents and
particle separation have been widely applied to study Lagrangian coherent
structures in oceanic flows. However, the influence of stable stratification on
Eulerian chaos and predictability in statistically stationary turbulence
remains largely unresolved.

In this work, we address this gap by studying uncertainty growth in forced,
statistically stationary three-dimensional stably stratified turbulence
governed by the Boussinesq equations. Using pseudo-spectral direct numerical
simulations, we perform twin simulations that differ by an infinitesimal,
localised perturbation at the initial time and track the evolution of velocity
and buoyancy differences across time and scales. This Eulerian decorrelator
framework enables the direct extraction of the largest Lyapunov exponent
$\lambda$ and a detailed characterisation of anisotropic uncertainty spectra as
a function of stratification strength. We show that increasing stratification
systematically suppresses chaoticity, primarily through a reduction of
nonlinear strain and enhanced strain--buoyancy alignment, rather than through
direct buoyancy effects. These results establish a quantitative link between
stratification, anisotropy and chaotic dynamics over a range of
Brunt--V\"ais\"al\"a frequencies relevant to oceanic conditions.

\section{Governing Equations, Numerical Methods, and Parameters}

Density-stratified turbulent flows require lifting the constant-density
approximation commonly employed for homogeneous and isotropic incompressible
turbulence. For stable stratification, the simplest background state consists
of a constant vertical density gradient $-\gamma$ imposed on a reference
density $\rho_0$ at $z=z_0$. In turbulent geophysical flows, however, density
fluctuations induced by motion must also be accounted for, so that the total
density field reads
\[
\rho(\mathbf{x},t)=\rho_0-\gamma(z-z_0)+\rho_f(\mathbf{x},t),
\]
where $\rho_f$ denotes the fluctuating component and gravity acts as
${\bf g}=-g\hat{\bf z}$. In the ideal, inviscid, and unforced limit, such a
background stratification supports small-amplitude vertical oscillations of
fluid parcels~\cite{Davidson_2013} with characteristic Brunt--V\"ais\"al\"a frequency
$N=\sqrt{g\gamma/\rho_0}$ .

When density fluctuations remain small compared to the background stratification,
the dynamics is well captured by the \textit{Boussinesq approximation}, in which
density variations affect the flow only through buoyancy forces. It is convenient
to work with the buoyancy field $b=(\rho_f N)/\gamma$, which has the same physical
dimensions as the velocity field. The governing equations for stably stratified
turbulence are then
\begin{eqnarray}
\frac{\partial {\bf u}}{\partial t}+{\bf u}\cdot\nabla{\bf u}
&=& \nu\nabla^2{\bf u}-\nabla p + N b\,\hat{\bf z}+{\bf f},\\
\frac{\partial b}{\partial t}+{\bf u}\cdot\nabla b
&=& \kappa\nabla^2 b - N\,{\bf u}\cdot\hat{\bf z},
\end{eqnarray}
supplemented by the incompressibility condition $\nabla\cdot{\bf u}=0$. The
external forcing ${\bf f}$ injects energy and maintains a statistically steady
turbulent state.

We solve these equations using pseudo-spectral direct numerical simulations
(DNSs) in a triply periodic domain of size $2\pi$, with up to $512^3$ collocation
points and a second-order Runge--Kutta time-marching scheme. The kinematic
viscosity is fixed at $\nu=5\times10^{-3}$, yielding Taylor-scale Reynolds
numbers
$100 \lesssim {\rm Re}_\lambda=u_{\rm rms}^2\sqrt{15/(\epsilon\nu)}\lesssim130$,
where $u_{\rm rms}$ is the root-mean-square velocity and $\epsilon$ is the mean
kinetic energy dissipation rate (the precise value depending on $N$). We set the
Prandtl number ${\rm Pr}=\nu/\kappa$ to unity, appropriate for thermally stratified
flows, though not representative of salt-stratified systems where ${\rm Pr}$
may be much larger. Simulations are performed for $N=0$ (the unstratified
reference case) and for $N=1,4,7,$ and $12$.

In this paper, we refer $\perp$ as the direction perpendicular to the axis (here, z-axis) of stratification. The flow is driven by a constant power injection applied only to the horizontal
velocity components (i.e., perpendicular direction). The forcing is vortical \cite{WAITE_BARTELLO_2004,Maffioli2016, Brethouwer2007,Varanasi2025} and restricted to modes with
$k_z=0$. In Fourier space, it is given by
\begin{equation}
\tilde{\boldsymbol{f}}_\perp(\mathbf{k}_\perp,k_z,t)=
\begin{cases}
\displaystyle
\frac{1}{M}\frac{\mathcal{P}}{2E^{(m)}_K}\,
\tilde{\mathbf{u}}_\perp(\mathbf{k}_\perp,k_z,t),
& (|\mathbf{k}_\perp|,k_z)=(m,0),\\[2mm]
0, & \text{otherwise},
\end{cases}
\end{equation}
with $\tilde{f}_z=0$. Here $M$ is the number of forced modes, $E^{(m)}_K$ is the horizontal
kinetic energy in the $(m,0)$ mode, and $\mathcal{P}$ is the prescribed power
input. In all runs, we force modes with $m=1,2$, so that $M=2$.

Although $N^{-1}$ defines the characteristic time scale associated with
stratification---analogous to the Kolmogorov time scale
$\tau_\eta=\sqrt{\nu/\epsilon}$---its dynamical influence is more conveniently
quantified by the Froude number
${\rm Fr}=u_{\rm rms}/(\ell N)$, where $\ell$ is the integral length scale
obtained from the kinetic energy spectrum $E_K(k)$. In our simulations,
$0.09\lesssim{\rm Fr}\lesssim0.97$, such that ${\rm Fr}<1$ and the buoyancy
Reynolds number ${\rm Re}_b\sim{\rm Re}_\lambda{\rm Fr}^2$ exceeds unity. Together
with ${\rm Re}_\lambda\gg1$, these conditions place our flows in the strongly
stratified turbulence regime.

To assess geophysical relevance, we compare our parameters with typical oceanic
values. Observations~\cite{Emery1984,Millard,techreport2021} indicate
$\mathcal{O}(10^{-3}\,\mathrm{s}^{-1})\lesssim N\lesssim
\mathcal{O}(10^{-1}\,\mathrm{s}^{-1})$. Using representative oceanic values
$\nu=10^{-6}\,\mathrm{m}^2/\mathrm{s}$~\cite{Stanley1969} and
$\epsilon=10^{-6}\,\mathrm{m}^2/\mathrm{s}^3$~\cite{Lien2021} yields
$\tau_\eta\simeq1\,\mathrm{s}$ and hence a dimensionless Brunt--V\"ais\"al\"a
frequency $\tilde N=N\tau_\eta$ in the range
$\mathcal{O}(10^{-3})\lesssim\tilde N\lesssim\mathcal{O}(10^{-1})$. This overlaps
with the values explored here, $0.082\lesssim\tilde N\lesssim1$ (with $N=\tilde N=0$
corresponding to the unstratified reference state).

\begin{figure}
\includegraphics[width=1.\linewidth]{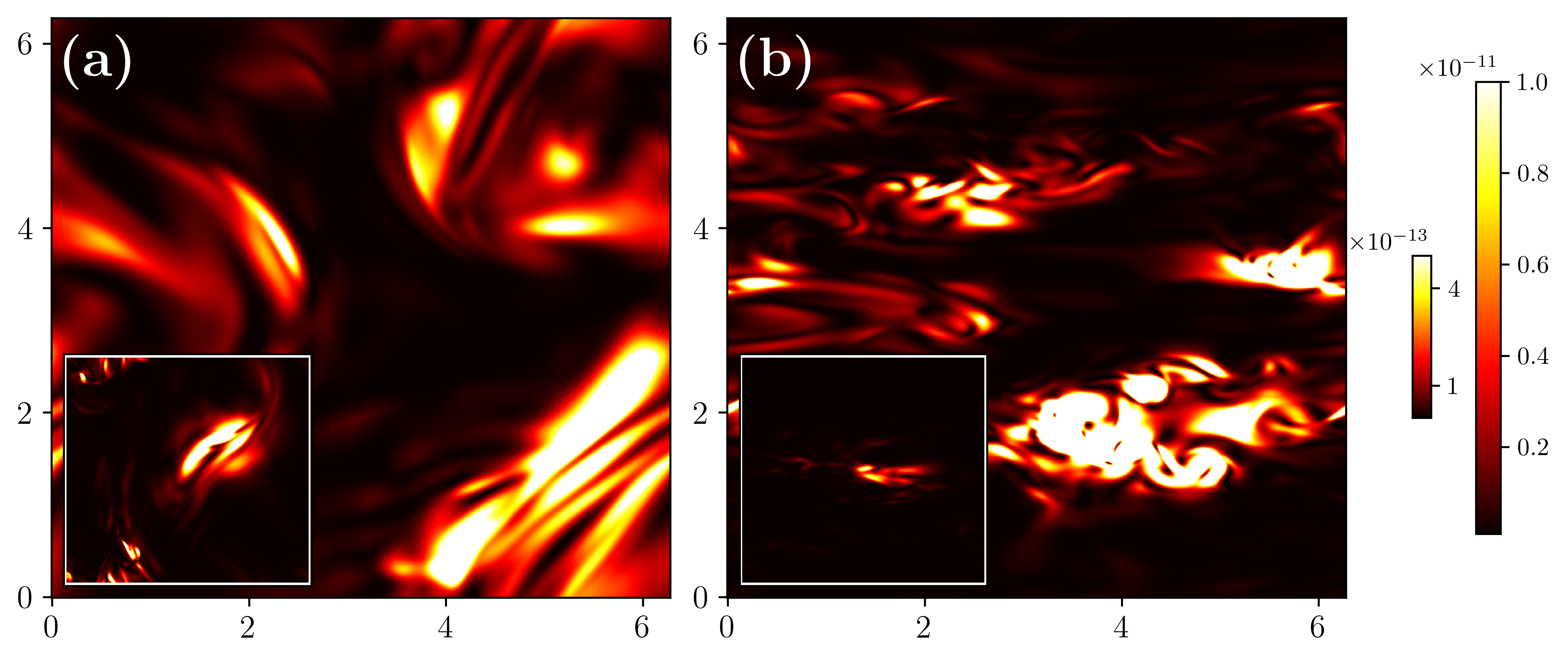}
\caption{Pseudocolor plots of 
$\phi_u(\mathbf{x},t)=\tfrac{1}{2}|\delta\mathbf{u}(\mathbf{x},t)|^2$
for a representative stratified run with $N = 12$ during the early stages of decorrelation 
	for slices in the (a) XY and (b) XZ planes. Insets correspond to an earlier time 
	before exponential
growth begins. The inner colorbar corresponds to the inset, while the outer colorbar corresponds to the main figure. The perturbation field develops pronounced anisotropy, forming
extended structures perpendicular to the direction of stratification. The
reduced vertical variability highlights the suppression of uncertainty
propagation along the stratification axis due to buoyancy effects.}
\label{fig:diff-field}
\end{figure}

Similarly, observed oceanic values of root-mean-square velocities
($0.1\,\mathrm{m/s}\lesssim u_{\rm rms}\lesssim2\,\mathrm{m/s}$) and integral length
scales ($\mathcal{O}(10^4)\,\mathrm{m}\lesssim\ell\lesssim
\mathcal{O}(10^5)\,\mathrm{m}$)~\cite{Jimenez} imply Froude numbers in the range
$\mathcal{O}(10^{-4})\lesssim{\rm Fr}\lesssim\mathcal{O}(10^{-2})$, consistent with
the regime probed by our simulations, albeit at lower Reynolds numbers than those
of the ocean.

\section{Uncertainty Growth and Lyapunov Exponents}

To quantify the sensitivity of stratified turbulence to infinitesimal
perturbations and thereby characterise uncertainty growth, we employ a twin-flow
construction in which two nearly identical states are evolved under identical
dynamics.

\begin{figure}
\centering
\includegraphics[width=1.0\textwidth]{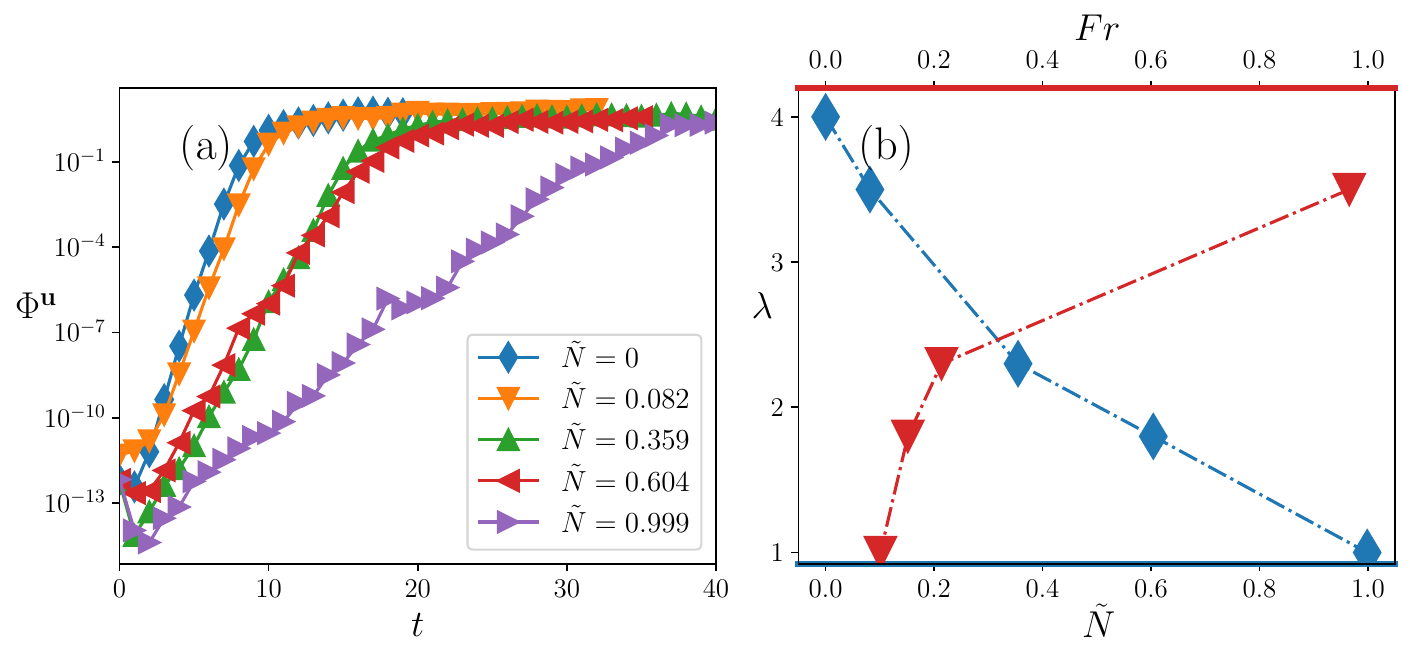}
\caption{(a) A semilog plot of the temporal evolution of the spatially averaged velocity decorrelator
$\Phi^u(t)=\langle \tfrac{1}{2}|\delta\mathbf{u}(\mathbf{x},t)|^2\rangle$
for different values of the non-dimensional Brunt--V\"ais\"al\"a frequency
	$\tilde N$ (see legend). After an initial decay, $\Phi^u(t)$ exhibits a clear exponential
growth regime followed by saturation, indicating complete decorrelation of
the two flow realizations. 
(b) The Lyapunov exponent $\lambda$ extracted from the exponential growth phase of
	$\Phi^u(t)$ as a function of $\tilde N$ (lower axis). The monotonic decrease of $\lambda$
with increasing stratification strength shows that stable stratification
systematically suppresses chaoticity. The same data is also plotted
	against the Froude number ${\rm Fr}$ (upper axis), confirming that increased buoyancy effects
lead to reduced sensitivity to infinitesimal perturbations.}
\label{fig:decorllyap}
\end{figure}

We begin by constructing statistically steady stratified turbulent states for
different values of the non-dimensional Brunt--V\"ais\"al\"a frequency $\tilde
N$. Let ${\bf u}^{\rm A}_0$ and $b^{\rm A}_0$ denote the velocity and buoyancy
fields of such a non-equilibrium steady state. A second, nearly identical flow
B is then initialised according to
\[
{\bf u}^{\rm B}_0 = {\bf u}^{\rm A}_0 + \epsilon_0 \nabla \times {\bf A},
\qquad
b^{\rm B}_0 = b^{\rm A}_0,
\]
where $\epsilon_0 \ll 1$.

The perturbation $\epsilon_0 \nabla \times {\bf A}$ is chosen to be small,
localized, symmetric, and divergence-free. Explicitly, the vector potential, $\mathbf{A}$ has
components
\[
A_i = \sqrt{\frac{3 u_{\rm rms}^2}{2}}\, r_0
\exp\!\left[-\frac{(\mathbf{x}-\mathbf{x}_0)^2}{2 r_0^2}\right]\hat{e}_i,
\]
with localization scale $r_0 \ll 2\pi$ and center at
$\mathbf{x}_0=(\pi,\pi,\pi)$. This choice ensures that the perturbation is
confined to a small region of the domain while preserving incompressibility.

Systems A and B are then evolved independently under identical dynamics. At any
later time $t$, this allows us to define the velocity and buoyancy
difference fields
\[
\delta {\bf u}(\mathbf{r},t)\equiv {\bf u}^{\rm B}-{\bf u}^{\rm A},
\qquad
\delta b(\mathbf{r},t)\equiv b^{\rm B}-b^{\rm A}.
\]
By construction,
$\delta{\bf u}(\mathbf{r},0)=\epsilon_0 \nabla\times{\bf A}$ and
$\delta b(\mathbf{r},0)=0$.

Fig.~\ref{fig:diff-field} shows representative pseudocolor plots of the
velocity decorrelator
$\phi^u(\mathbf{x},t)=\tfrac{1}{2}|\delta\mathbf{u}(\mathbf{x},t)|^2$
for a typical stratified run at an early time $t = 7$ during the decorrelation process.
We display two-dimensional slices taken in the (a) XY and (b) XZ planes. 
The insets correspond to an earlier instant $t = 2$, before
exponential growth sets in. These visualisations reveal the emergence of
strongly anisotropic structures, with the perturbation field exhibiting
quasi-two-dimensional organisation perpendicular to the direction of
stratification. In particular, our results indicate an inhibited spread of uncertainty in $z$ due
to buoyancy effects.

To quantify uncertainty growth, we consider the spatially averaged decorrelator
\[
\Phi^u(t)=\langle \phi^u(\mathbf{x},t)\rangle ,
\]
where angular brackets denote a volume average. Fig.~\ref{fig:decorllyap}(a) shows the temporal
evolution of $\Phi^u(t)$ for several values of the non-dimensional
Brunt--V\"ais\"al\"a frequency, $\tilde N$. For all cases, $\Phi^u(t)$ displays a
common sequence of dynamical regimes. At very early times, $\Phi^u(t)$ exhibits
a slight decay, reflecting the rapid viscous relaxation of the imposed
perturbation. This is followed by a pronounced growth phase in which
$\Phi^u(t)$ increases exponentially in time before eventually saturating once
the two flow realizations become fully decorrelated.

The existence of a clear exponential growth regime allows us to define a
Lyapunov exponent $\lambda$ characterising the rate of uncertainty
amplification. Extracting $\lambda$ from the slope of $\ln \Phi^u(t)$ during
this interval, we find that the degree of chaoticity decreases systematically as
stratification strengthens. As shown in Fig.~\ref{fig:decorllyap}(b), $\lambda$ decreases
monotonically with increasing $\tilde N$, indicating that stable stratification
suppresses sensitivity to infinitesimal perturbations. The same trend is evident in Fig.~\ref{fig:decorllyap}(b) 
when $\lambda$ is plotted against the Froude number, confirming that enhanced
buoyancy effects lead to a dynamically calmer regime compared to homogeneous and
isotropic turbulence at comparable Reynolds numbers.

\section{Physical Interpretation of Uncertainty Growth}

How can these observations be understood directly from the equations of motion?
Following the framework developed in
Refs.~\cite{banerjee2025intermittent,Vassilicos_2023}, we derive an evolution
equation for the spatially averaged velocity decorrelator $\Phi^u(t)$:
\begin{equation}
\label{rate_equation}
\partial_t \Phi^u(t) =
\beta_S + \beta_\eta + N\langle \delta u_z\,\delta b\rangle
+ \langle \delta\mathbf{f}\cdot\delta\mathbf{u}\rangle ,
\end{equation}
where
$\beta_S=-\langle \delta\mathbf{u}\cdot\mathbf{S}\cdot\delta\mathbf{u}\rangle$
represents stretching and compression by the symmetric rate-of-strain tensor
$\mathbf{S}$, and
$\beta_\eta=\nu\langle \delta\mathbf{u}\cdot\nabla^2\delta\mathbf{u}\rangle$
denotes viscous dissipation. In addition to the terms familiar from homogeneous
and isotropic turbulence (HIT), stratification introduces the buoyancy–velocity
cross-correlation $N\langle \delta u_z\,\delta b\rangle$.

\begin{figure}
\includegraphics[width=1.0\linewidth]{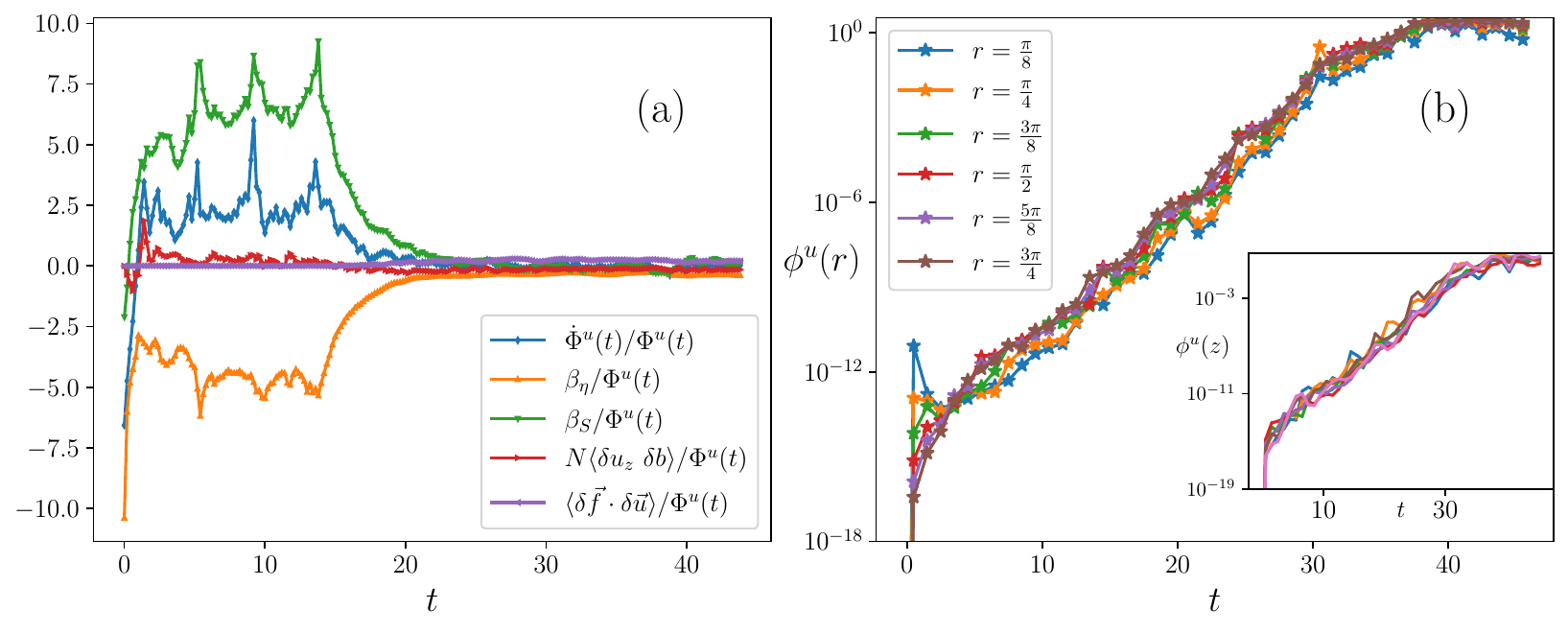}
\caption{(a) Time evolution of the individual contributions to the decorrelator growth
rate in Eq.~\eqref{rate_equation} for $N = 4$: the strain-induced term $\beta_S$,
the viscous term $\beta_\eta$, the buoyancy–velocity cross-correlation
$N\langle\delta u_z\,\delta b\rangle$, and the forcing contribution, each
normalized by $\Phi^u(t)$. The initial decay of $\Phi^u(t)$ is driven by large,
negative contributions from $\beta_S$ and $\beta_\eta$, while the buoyancy term
remains subdominant throughout.
(b) Temporal evolution of the spherically averaged decorrelator
$\phi^u(r,t)$ with respect to the centre at $\mathbf{x}_0$ for different radii $r$ (see legend), illustrating spatially homogeneous
exponential growth of uncertainty. The inset shows $\phi^u(z,t)$ at fixed
$r$ for seven different values of $z$ in the range $0.53 \pi \leq z \leq 1.47 \pi$ , confirming uniform growth even along the direction of stratification. The dashed curve represents the spectra for the reference state. }
\label{fig:rate_and_decorwitrad}
\end{figure}

Direct evaluation of these contributions shows that, throughout most of the
evolution and across all Froude numbers considered, the buoyancy-induced term
remains subdominant compared to $\beta_S$ and $\beta_\eta$, as illustrated in
Fig.~\ref{fig:rate_and_decorwitrad}(a). Immediately after the perturbation is
introduced, both $\beta_S/\Phi^u(t)$ and $\beta_\eta/\Phi^u(t)$ are large and
negative, resulting in the initial decay of $\Phi^u(t)$ observed in
Fig.~\ref{fig:decorllyap}(a). This transient reflects the rapid viscous relaxation
of a perturbation that has not yet developed preferential alignment with local
flow structures.

Further insight is obtained by decomposing $\beta_S$ in the eigenbasis of the
rate-of-strain tensor,
\[
\beta_S = -\sum_{i=1}^3 n_i^2\,\gamma_i\,|\delta\mathbf{u}|^2,
\]
where $\gamma_i$ are the eigenvalues of $\mathbf{S}$ and $n_i$ are the direction
cosines of $\delta\mathbf{u}$ along the corresponding eigendirections.
The largest eigenvalue $\gamma_1>0$ corresponds to the extensional direction,
$\gamma_3<0$ to the compressional direction, while $\gamma_2\approx 0$ is
associated with a nearly neutral direction. At early times, the perturbation
samples these eigendirections almost isotropically, so that
$\sum_i n_i^2\gamma_i$ is negative on average and $\beta_S$ contributes to decay.
During the exponential growth phase, however, the perturbation aligns strongly
with the compressional eigendirection ($n_3^2\simeq 1$), while the
contribution from extensional stretching gets suppressed. In this regime, $\beta_S+\beta_\eta$
becomes positive, driving exponential uncertainty growth, while both the
buoyancy–velocity correlator and forcing term remain negligible.

A further hallmark of incompressible Navier–Stokes dynamics is the non-locality
of the pressure field, which implies that an imposed perturbation can influence
distant regions of the flow essentially instantaneously. As a result, the
growth of uncertainty is spatially homogeneous. This behaviour is
demonstrated in Fig.~\ref{fig:rate_and_decorwitrad}(b), where we show the temporal
evolution of the spherically averaged decorrelator
$\phi^u(r,t)=\langle\phi^u(\mathbf{x},t)\rangle$ for several values of the radius, $r = \vert \mathbf{x} - \mathbf{x}_0 \vert$ with respect to the center, $\mathbf{x}_0$ of the initial perturbation. All radial shells exhibit identical exponential
growth rates. One may go further and define $\phi^u(z,t)$ as the azimuthally averaged decorrelator with respect to the z-axis passing through $\mathbf{x}_0$.  The inset of Fig.~\ref{fig:rate_and_decorwitrad}(b) shows the evolution of $\phi^u(z,t)$ for various values of $z$ in the shell defined by $r = \pi/2$. This further confirms this homogeneity by demonstrating that uncertainty growth remains uniform also along
z direction despite the underlying anisotropy of stratified turbulence.

\section{Spectral and Anisotropic Structure of Uncertainty}

\begin{figure}
\includegraphics[width=1.0\textwidth]{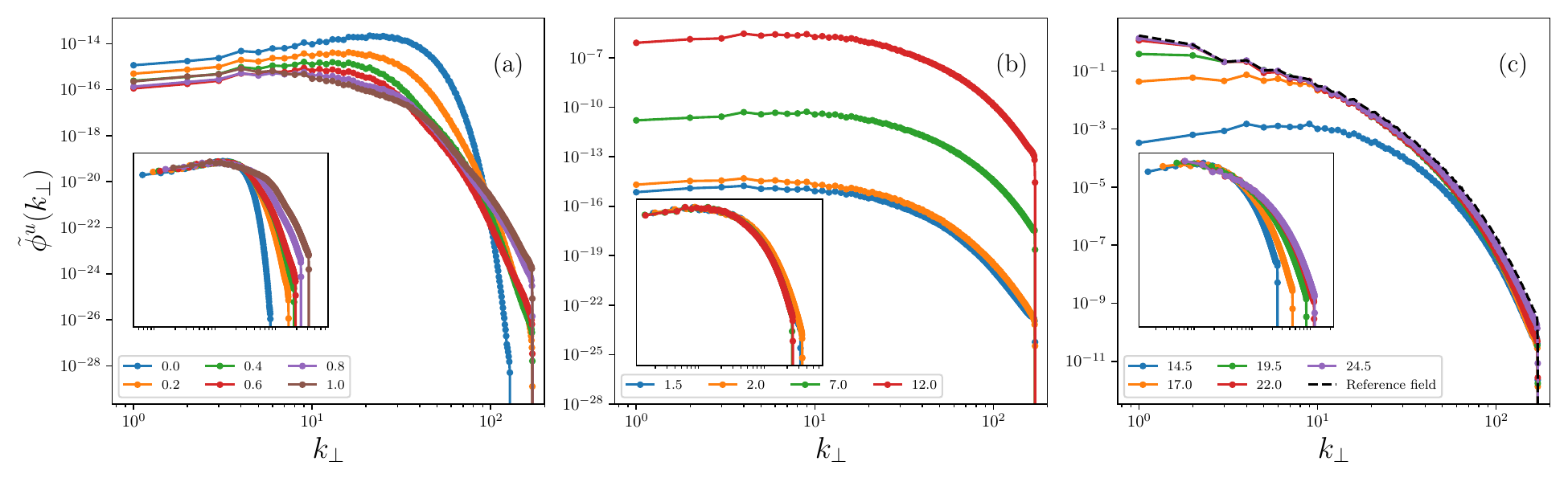}\\
\includegraphics[width=1.0\textwidth]{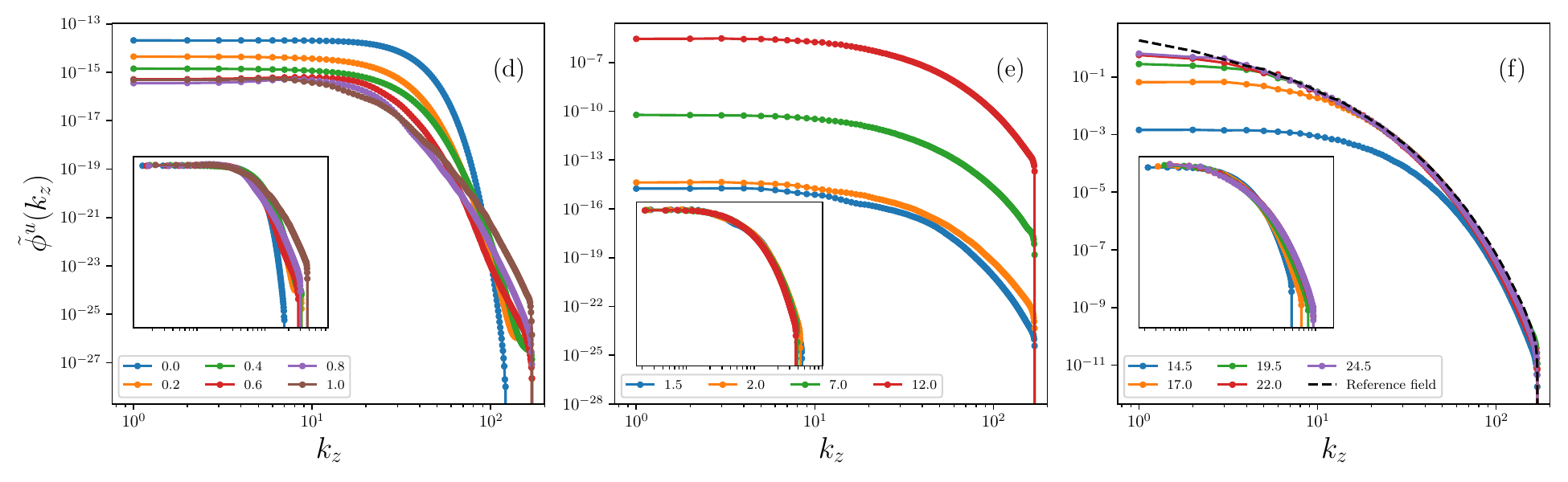}
\caption{Evolution of the uncertainty spectrum in spectral space:
(a)--(c) Horizontal decorrelator spectra $\tilde{\phi}^u(k_\perp,t)$ at
representative times during the decay, exponential growth, and saturation
phases. The inset shows the rescaled spectra
$\tilde{\phi}^u(k_\perp,t)/[\ell^\Delta_\perp(t)\Phi^u(t)]$
plotted against $k_\perp \ell^\Delta_\perp(t)$. As can be seen from inset of (b), the rescaled spectra collapses 
demonstrating self-similar evolution in the growth phase.
(d)--(f) The corresponding vertical decorrelator spectra
$\tilde{\phi}^u(k_z,t)$, exhibiting analogous behaviour along the direction of
stratification.}
\label{fig:kz}
\end{figure}

The anisotropic nature of uncertainty growth, evident in the real-space
decorrelator fields shown in Fig.~\ref{fig:diff-field}, can be quantified more
systematically in spectral space. Unlike homogeneous isotropic turbulence,
stably stratified flows possess an inherent direction set by gravity, and the
spread of uncertainty along the vertical ($z$) direction differs markedly from
that in the horizontal plane. This motivates a scale-dependent analysis of the
decorrelator using uncertainty spectra.

\begin{figure}
\includegraphics[width=0.9\linewidth]{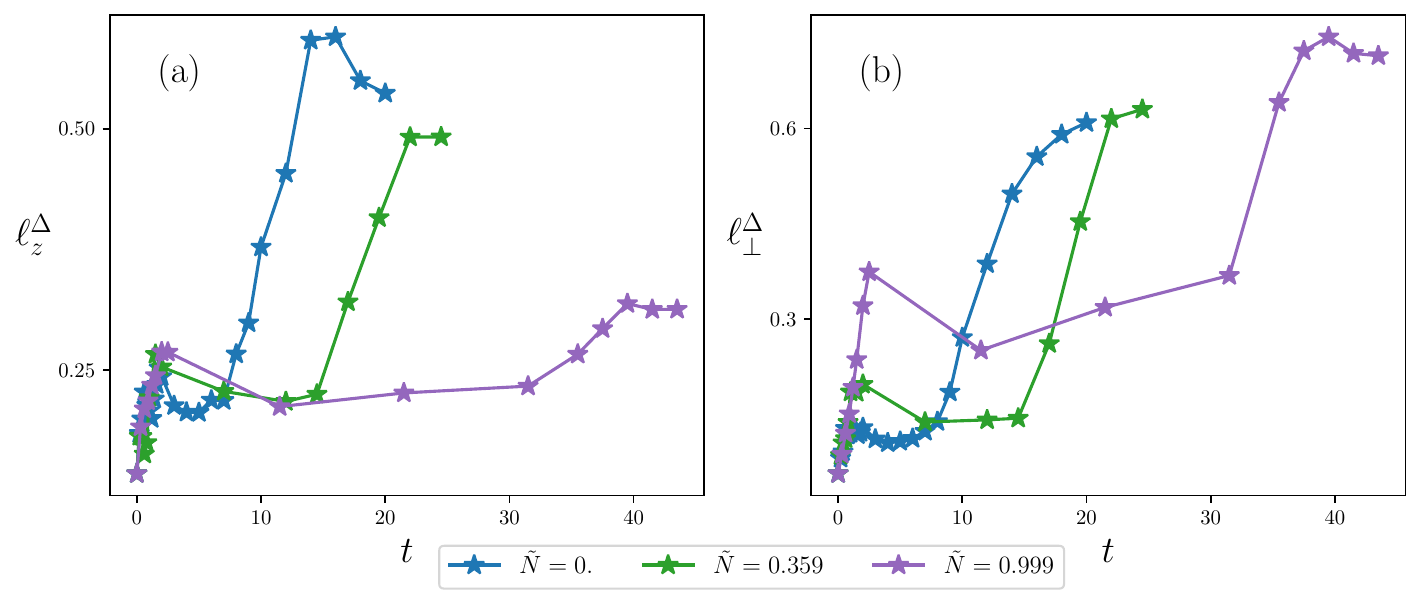}
\caption{Temporal evolution of the uncertainty integral length scales in the (a) vertical
($\ell^\Delta_z$) and (b) horizontal ($\ell^\Delta_\perp$) directions. The evolution is shown till the time the two systems A and B decorrelate completely. Throughout the
exponential growth phase, the vertical uncertainty scale remains significantly
smaller than the horizontal one, reflecting the anisotropic suppression of
vertical motions by stable stratification.}
\label{fig:Integral_scale_with_time}
\end{figure}

We define the velocity decorrelator spectrum as
\begin{equation}
\tilde{\phi}^u(\mathbf{k},t)
= \frac{1}{2}\left\langle
\left|\delta\tilde{\mathbf{u}}(\mathbf{k},t)\right|^2
\right\rangle ,
\end{equation}
where $\delta\tilde{\mathbf{u}}(\mathbf{k},t)$ denotes the three-dimensional
Fourier transform of $\delta\mathbf{u}(\mathbf{x},t)$, and
$\langle\cdots\rangle$ represents an average over the spherical shell
$|\mathbf{k}|=\text{const}$. This yields the isotropic uncertainty spectrum.
To probe anisotropy, we additionally consider reduced spectra,
$\tilde{\phi}^u(k_\perp,t)$ and $\tilde{\phi}^u(k_z,t)$, obtained by summing
over wavevectors with fixed $k_\perp=\sqrt{k_x^2+k_y^2}$ and fixed $k_z$,
respectively.

Fig.~\ref{fig:kz}(a)--(c) shows the temporal evolution of
$\tilde{\phi}^u(k_\perp,t)$ across distinct phases of decorrelation. During the
initial decay phase, uncertainty rapidly populates higher horizontal
wavenumbers. This is followed by a clear exponential-growth regime in which the
spectrum evolves self-similarly. In this regime, the decorrelator spectrum
exhibits the scaling form
\begin{equation}
\tilde{\phi}^u(k_\perp,t)
\sim
\ell^\Delta_\perp(t)\,\Phi^u(t)\,
f\!\left(k_\perp \ell^\Delta_\perp(t)\right),
\end{equation}
where $f$ is a universal scaling function and
$\ell^\Delta_\perp(t)$ is the uncertainty integral length scale in the
perpendicular direction.

The integral scale is defined as
\begin{equation}
\ell^\Delta_\perp(t)
=
\frac{1}{\sum_{k_\perp}\tilde{\phi}^u(k_\perp,t)}
\sum_{k_\perp}
\frac{\tilde{\phi}^u(k_\perp,t)}{k_\perp},
\end{equation}
and measures the dominant horizontal scale over which uncertainty is
distributed. The inset of Fig.~\ref{fig:kz}(b) demonstrates this
self-similarity: when the spectra are rescaled by $\ell^\Delta_\perp(t)$ and
$\Phi^u(t)$, they collapse onto a single curve during the exponential growth
phase.

As the decorrelator approaches saturation, uncertainty progressively invades
lower wavenumbers, indicating a transfer from small to large scales. Eventually,
once the two realizations are completely decorrelated, the uncertainty spectrum
converges to the kinetic energy spectrum of the reference flow,
\begin{equation}
\tilde{\phi}^u(k_\perp,t) \;\longrightarrow\; 2E_K(k_\perp),
\end{equation}
shown by dashed black curves in Fig.~\ref{fig:kz}(c). An analogous sequence of
spectral phases is observed for the vertical spectrum
$\tilde{\phi}^u(k_z,t)$, shown in Fig.~\ref{fig:kz}(d)--(f), as well as for the
isotropic uncertainty spectrum (not shown). Similar self-similar evolution of
decorrelator spectra was previously reported for homogeneous isotropic
turbulence~\cite{Vassilicos_2023}.

Strong stratification is known to suppress vertical velocity fluctuations,
leading to flow structures with much smaller vertical length scales than
horizontal ones. This anisotropy is reflected directly in the uncertainty
dynamics. As seen in Fig.~\ref{fig:diff-field}, the decorrelator structures in
the growth phase are elongated in the horizontal plane and compressed along
$z$. This behaviour is quantified in Fig.~\ref{fig:Integral_scale_with_time},
which compares the temporal evolution of the uncertainty integral length scales
in the horizontal ($\ell^\Delta_\perp$) and vertical ($\ell^\Delta_z$)
directions. Throughout the exponential growth phase,
$\ell^\Delta_z \ll \ell^\Delta_\perp$, with the disparity increasing as the
stratification strength is raised.

Finally, an analogous decorrelator $\phi^b(\mathbf{x},t)$ may be defined for the
buoyancy field. We find that the spatially averaged buoyancy decorrelator
$\Phi^b(t)$ grows at the same exponential rate as its velocity counterpart
$\Phi^u(t)$. Consequently, buoyancy fluctuations exhibit spectral evolution and
anisotropy qualitatively similar to those observed for the velocity field.

\section{Conclusions}

We have investigated the growth of uncertainty in stably stratified
turbulence using decorrelator-based diagnostics and Lyapunov analysis.
Increasing stratification strength (decreasing Froude number) leads to a
systematic reduction of the Lyapunov exponent, demonstrating that stable
stratification suppresses chaoticity. Despite this quantitative change,
uncertainty growth retains the familiar temporal structure observed in
homogeneous isotropic turbulence: an initial decay phase, followed by
self-similar exponential growth and eventual saturation.

The exponential growth regime is characterised by a universal, self-similar
evolution of the uncertainty spectra. However, in contrast to isotropic flows,
uncertainty propagation is strongly anisotropic. Perturbations spread much more
slowly in the vertical direction than in the horizontal plane, and the
uncertainty integral scale along the stratification direction remains
consistently smaller throughout the growth phase. This anisotropy intensifies
with increasing stratification, reflecting the inhibition of vertical motions
by buoyancy forces.

Analysis of the decorrelator evolution equation reveals the physical origin of
this reduced chaoticity. Stratification primarily acts by altering the
alignment between perturbations and the local rate-of-strain tensor. During the
exponential growth phase, perturbations preferentially align with compressive
eigendirections, diminishing the effectiveness of extensional stretching that
drives rapid uncertainty amplification in isotropic turbulence. In contrast,
buoyancy--velocity cross-correlations and forcing contributions remain
subdominant, indicating that the suppression of chaos is controlled
predominantly by strain-mediated dynamics rather than direct buoyancy coupling.
Although the Reynolds numbers achieved in our simulations remain lower than
those characteristic of oceanic flows, similar limitations are common in direct
numerical studies of stratified turbulence. Nevertheless, the systematic trends
reported here --- particularly the monotonic suppression of the Lyapunov
exponent with increasing stratification and the anisotropic structure of
uncertainty growth --- are expected to persist at higher Reynolds numbers.

From a geophysical perspective, these findings suggest that strong stable
stratification can enhance short-time predictability by limiting the vertical
spread of uncertainty, even while permitting rapid horizontal error growth.
Taken together, our results provide a quantitative characterization of
uncertainty growth and Lyapunov dynamics in stratified turbulence. Beyond their
implications for predictability in geophysical flows, this work establishes
decorrelator-based methods as a robust framework for diagnosing chaotic
dynamics in anisotropic and wave-supporting turbulent systems, with natural
extensions to rotating, magnetized, and convectively driven flows.

\begin{acknowledgments} 
 MJP and SSR acknowledge several discussions with Arun Kumar Varanasi. MJP also thanks Rajarshi and Sanjay C P 
	 for useful suggestions.
	SSR acknowledges the Indo–French Centre for the Promotion of Advanced Scientific Research
	(IFCPAR/CEFIPRA, project no. 6704-1) for support. The simulations were performed on the
	ICTS clusters Mario, Tetris, Boson2, and Contra. MJP and SSR acknowledge the
	support of the DAE, Government of India, under projects nos.
	12-R\&D-TFR-5.10-1100 and RTI4001. 
\end{acknowledgments}

\bibliography{references}
\end{document}